\documentclass[a4paper,journal,doublecolumn,10pt]{IEEEtran}
\usepackage{graphicx}
\usepackage{dirtree}
\usepackage{caption}
\usepackage{soul}
\usepackage{xcolor}
\usepackage{cite}
\usepackage{multirow}
\usepackage{nicematrix}

\IEEEoverridecommandlockouts

\begin{document}

\title{
 AI Testing Framework for Next-G O-RAN Networks: Requirements, Design, and Research Opportunities}

\author{Bo Tang,~\textit{Senior, IEEE,} \thanks{Bo Tang and Vuk Marojevic are with the Electrical and Computer Engineering Department at Mississippi State University. Vijay K. Shah is with Cybersecurity Engineering Department at George Mason University. Jeffrey H. Reed is with the Department of Electrical and Computer Engineering at Virginia Tech. Emails: \{tang, vuk.marojevic\}@ece.msstate.edu, vshah22@gmu.edu, reedjh@vt.edu. This research is supported by the National Science Foundation, under grant number $2120411$ and $2120442$. }Vijay K. Shah, \textit{Member, IEEE}, Vuk Marojevic, \textit{Senior, IEEE} and Jeffrey H. Reed, \textit{Fellow, IEEE}}

\maketitle

\begin{abstract}
Openness and intelligence are two enabling features to be introduced in next generation wireless networks, e.g. {Beyond 5G} and 6G, to support service heterogeneity, open hardware, optimal resource utilization, and on-demand service deployment. The open radio access network (O-RAN) is a promising RAN architecture to achieve both openness and intelligence through virtualized network elements and well-defined interfaces. While deploying artificial intelligence (AI) models is becoming easier in O-RAN, one significant challenge that has been long neglected is the comprehensive testing of their performance {in} realistic environments. This article presents a general automated, distributed and AI-enabled testing framework to test AI models deployed in O-RAN in terms of their decision-making performance, vulnerability and security. This framework adopts a master-actor architecture to manage a number of end devices for distributed testing. More importantly, it leverages AI to automatically and intelligently explore the decision space of AI models in O-RAN. Both software simulation testing and software-defined radio hardware testing are supported, {enabling rapid proof of concept research and experimental research on wireless research platforms.} 
\end{abstract}


\IEEEpeerreviewmaketitle

\section{Introduction}
\label{sec:intro}

Cellular communications networks have evolved 
from an inflexible and monolithic system to a flexible, agile and disaggregated architecture. 
Leveraging research innovations, next generation (Next-G) networks are expected to be built and operated based on the \textit{openness} and \textit{intelligence} principles. 
Furthermore, Beyond 5G and future 6G networks will incorporate artificial intelligence (AI) into the deployment, operation, and maintenance of the network \cite{wang2020artificial}
\cite{shafin2020artificial}. AI is well suited for communications; it is useful for estimating near-optimal settings in situations that have a large search space, 
it can generalize a solution to respond to new situations, it can optimize the network's operation when resources are limited, 
and it can interpolate when insufficient information is available. 

Modern networks can be classified by the degree of integration of AI. {The European Telecommunications Standards Institute (ETSI)} defines six stages for network automation based on the use of AI, from no AI (i.e., manual control) to fully AI-driven systems (i.e., cognitive AI) \cite{etsi}.  
The {Global System for Mobile Communications Association (GSMA)} supported the development of two major applied AI initiatives in 2019, aimed at sharing insight and developing an expert community: Applied AI Forum and the GSMA Global AI Challenge. 
The open radio access network (O-RAN) is an industry-driven architecture with open interfaces and open-source implementations \cite{ORANwhitepaper}. 
The O-RAN Alliance is considering AI as an integral part of its open architecture and RAN control framework. 

While AI models are enablers to achieve intelligent Next-G wireless networks, comprehensive testing of their performance is cumbersome and in many cases non-existent. This is mainly due to the inability of the current theory to explain or prevent failures in the AI models. Hence, it is necessary to have a framework and appropriate environment for testing AI models in their capacity of 
cellular RAN controllers. From the ongoing 
research and development and expected deployment of O-RAN components in Next-G networks, there is {an 
urgent} need for methods, platforms, and tools that facilitate testing various AI models in the radio network in a production like environment \cite{oaic_msstate}. 

The Third Generation Partnership Project (3GPP) standardizes cellular communications. It defines the architecture, protocols, parameters, signals, and so forth. It also defines test procedures and expected outcomes for building 3GPP compliant networks. These can be generally categorized as performance and compliance tests, which are related to signal processing and {radio frequency (RF)} transmission, among others. 
The O-RAN specifications provide different options where and how to train the AI models, but not how to test the operations of the RAN Intelligent Controllers (RICs). 

This paper presents a testing framework design\footnote{{https://github.com/openaicellular}} to help optimize the evaluation of {AI controlled O-RAN systems} under variations of the input data in realistic and, possibly, changing conditions. This framework supports automated and distributed testing by managing a number of test actors which are able to transmit testing signals in parallel. Given the large input space of an AI-controlled system, it may be impossible to exhaustively and comprehensively test {the} performance of AI models without the help of AI. Our framework integrates AI-enabled testing methods to explore the decision space of AI models as cellular RAN controllers.

\section{Cellular Network Testing Standards 
and Research}
\label{sec:etsi}

\subsection{Cellular Network Testing Standards}
Several standardization bodies work on establishing standards for Next-G cellular network testing, including 3GPP, O-RAN Alliance and ETSI.

\noindent \textbf{3GPP:} {Testing procedures are} 
outlined in the 3GPP specifications which cover functional, performance, and conformance tests by establishing RF transmission masks, signaling requirements, and expected performance figures, among others. 
{These procedures} 
are adopted by network and device manufacturers using purpose-built test instruments and test user equipment (UEs), and {are leveraged} by other stakeholders, including the O-RAN Alliance, providing new features and services on top of 3GPP protocols.

\noindent \textbf{O-RAN Alliance:} 
The O-RAN Alliance is a consortium that develops standards for open cellular networks. It was formed by merging the C-RAN Alliance 
and the xRAN Forum. 
Its 
mission is to extend the current RAN standards and facilitate open, intelligent, virtualized, and fully interoperable Next-G RANs~\cite{ORANwhitepaper}. 
The O-RAN Alliance extends the 3GPP standards with new interfaces and intelligent {controllers}, creating a framework for developing and deploying intelligent, software-defined networking (SDN) based, and virtualized cellular networks while leveraging 3GPP's 4G and 5G protocols and network components, encompassing the RAN, UE and core network. 

Similar to the 3GPP, the O-RAN Alliance's WG4 defines test cases, parameters, and procedures for testing the conformance and performance of the {O-RAN distributed unit (O-RU), control unit (O-CU), and radio unit (O-RU)}. 
The O-RAN Alliance Test and Integration {Focus Group (FG)}, furthermore, recently published specifications 
{that} establish the scope, goals, and processes for end-to-end network testing, where the O-RAN system under test is treated as a \textit{black box}~\cite{ORAN_Testing}. {These specification cover} functional, performance,  service, security, load and stress tests. {They do} not specify how to evaluate the AI models or network intelligence. 

In addition to test procedures, the O-RAN Alliance has established O-RAN Test and Integration Centers (OTICs), which are independent test platforms/sites where vendors can test their O-RAN systems. This is in an early stage and vendor-neutral institutes are encouraged to apply and become an OTIC, following the O-RAN guidelines. 
The Test and Integration FG also establishes certification and badging processes and procedures to be used by OTICs to certify vendor products.

\noindent \textbf{ETSI:}
A work item on \textit{AI in testing systems and testing AI models} has been published by ETSI \cite{etsi_testing}. It is publicly available and introduced in their 5G Proof of Concept (PoC) White Paper \#5. This is part of ETSI's Generic Autonomic Network Architecture program and defines a generic 
framework for testing AI models/systems, from the 
validation of AI model to network optimization. 
It includes data, {algorithm} and model validation, as well as non-functional and integration testing. 

ETSI and its stakeholders 
{conceptualize} an offline 
training and test environment and envisage it to {interface} with production networks for obtaining network data and providing non-real time feedback.
They introduce 
slow and fast control loops as part of the production network \textit{knowledge plane} and the production network itself. The PoC White Paper~\#5 proposes components and process flows for testing AI systems/networks, without providing specifics about the networks, management tasks, or AI models.
While ETSI {emphasizes} 
the need for AI system testing and {defines} a general framework and sample processes, it does not specify how to test and verify the RAN specific interfaces, AI models, individual or compound network functions, or RAN controllers.



\subsection{RAN Testing Research, Methods, and Technologies
}
Research has shown deficiencies in the 4G and 5G wireless protocols by means of non-standard and innovative testing in controlled laboratory environments. 
For example, security vulnerabilities of 4G and 5G networks have been discovered by jamming, spoofing, eavesdropping, and other types of systematic attacks applied on network modules or interfaces using 
{software-defined radios (SDRs)}. 
{SDR hardware and open-source software 
facilitate demonstrating specific wireless protocol vulnerabilities 
and implementing and evaluating fixes 
for these}~\cite{mina17}. Systematic radio attacks to {a commercial 4G radio access network} for mission critical applications have been demonstrated in~\cite{sean17}. The paper shows loss in system performance with targeted radio interference, implemented in software and transmitted from {SDRs}. It also proposes {machine learning (ML) techniques} to process performance measurement counters and key performance indicators {(KPIs)} collected by the network to detect and classify 
attacks. 

Byrd et al.~\cite{Byrd2020} introduce {an} open-source cellular 
security analysis instrument which provides a practical tool to observe and analyze control messages between the cell towers and UEs. It uses open-source software and commercial off-the-shelf (COTS) {SDR} hardware and can, for instance, observe the use and lifetime of Temporary Mobile Subscriber Identities that are meant to be used as temporary identifiers to authenticate users and protect their identities. 
ProChecker~\cite{prochecker} combines dynamic testing with static instrumentation to extract a semantic model of 4G protocol interactions. 
{It uses} a symbolic model checker {together with} a cryptographic protocol verifier to verify the properties against the extracted model 
and to analyze 4G control plane protocol implementations against a variety of security attacks. 


\section{
The O-RAN Context and  Requirements}
\label{sec:oran}
\subsection{The O-RAN Architecture}



\begin{figure}
    \centering
    \includegraphics[scale=0.38]{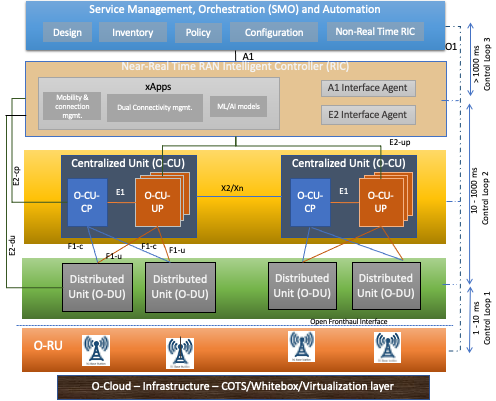}
    \caption{O-RAN Architecture}
    \label{fig:o-ran}
\end{figure}

{O-RAN is an emerging, transformational RAN architecture 
that emphasizes openness, intelligence, virtualization, softwarization, disaggregation, and multi-vendor support.} This can provide several benefits, such as, reduced cost of maintenance, dynamic services, quicker time to market for new user and network management services as well as other innovations 
transforming the telecom industry. O-RAN adopts 
{functional splits 2 and 7.2x} 
to disaggregate base station functionalities into the O-CU, O-DU, and O-RU. Whereas, {the O-RU may be implemented in 
hardware}, O-RAN supports software-based O-DU and O-CU {implementations that may be} hosted as virtual network functions on 
{edge or cloud servers}. These RAN components are connected via open interfaces standardized by 3GPP {and the O-RAN Alliance}. 

{The O-RAN architecture {features}  two logical controllers to facilitate the management, control, and orchestration of the network with closed-loop control. The \textit{near-real time RIC} ({near-RT RIC}) is deployed at the edge of the network and operates on a timescale between $10$ ms and $1$ s. It establishes the means for 
monitoring, management, control, and orchestration of the O-DUs and O-CUs in the RAN. 
The \textit{non-real time RIC} ({non-RT RIC}) operates at a time scale of $1$~s and above. It facilitates the 
orchestration of network resources at the infrastructure level 
through 
policies that may impact network operations and users. 
O-RAN therefore establishes 
additional open 
interfaces: A1 between the non-RT RIC and near-RT RIC and E2 between the near-RT RIC and the O-CU/O-DU. 
}

{The near-RT RIC can host multiple 
\textit{xApps} and the required services to manage their life cycle. An xApp is a microservice that 
may collect data from the RAN (e.g., user and cell key performance measurements
, such as number of users, load, throughput, resource utilization), process the data, and 
send back control actions to the RAN through standardized interfaces. 
Examples of xApps for near-RT RAN control applications 
include 
scheduling, traffic shaping, and handover management.}

{The non-RT RIC supports the execution of 
\textit{rApps}, which 
facilitate RAN optimization and operation, including policy guidance, enrichment information, and configuration management. Although rApps can support the same RAN control functionalities provided by xApps (e.g., traffic shaping, 
scheduling, and handover management) at larger timescales
, they have been standardized to derive control and management policies that operate at a higher level and affect a large number of RAN nodes and users. 
Examples of rApps for non-RT RAN control applications include frequency management, network slicing, and policy management. 
}



{Both xApps and rApps may be data-driven and employ AI/ML. 
The O-RAN Alliance defines different deployment options for the training and inference of such AI controllers \cite{ORANTechRep20}. 
The 
non-RT RIC hosts the ML training and the non-RT RIC or near-RT RIC can host the ML inference.}

\subsection{AI Integration in O-RAN}

The O-RAN architecture enables the integration of AI models to perform intelligence decisions based on {network} and environmental conditions. 
The use of AI, leveraging the collected information, helps to enhance both cellular performance per cell and user performance per UE, such as long-term traffic congestion, latency, cell coverage, radio interference, among many other KPIs. This can be achieved by deploying a number of AI models including, but not limited to, supervised learning, unsupervised learning, reinforcement learning, federated learning, and transfer learning~\cite{ORANTechRep20}. We summarize possible AI models to be integrated in O-RAN for different applications in Table \ref{tab:AI}.

\begin{table*}[h!]
\centering

\caption{Integration of various AI models in O-RAN.}
\begin{tabular}{ |c|p{8cm}|c|c|  }
 \hline
 Type of AI & Application 
 & Network Layer & RIC\\
 \hline
 \multirow{3}{*}{Supervised Learning}   & Prediction of traffic, e.g., type and volume, using regression algorithms. & MAC &  Near-RT RIC \\ \cline{2-4}
 & Prediction of available bandwidth at different times and locations using regression algorithms. & MAC & Near-RT RIC\\\cline{2-4}
 & Identification of a set of predefined modulations using classification algorithms to quickly  classify the modulation type of the interference signal.  & PHY & RT RIC\\\cline{2-4}
 & Identification of a set of certain intrusions using classification algorithms to detect network born attacks such as denial of service attacks, flooding, and twin-evil intruders.  & MAC & Near-RT RIC\\
 \hline
  \multirow{3}{*}{Unsupervised Learning}   & Detection of abnormal traffic using anomaly detection algorithms to identify potential attacks. & MAC & Near-RT RIC  \\ \cline{2-4}
 & Clustering of traffic, e.g., type and demand, to identify similar interests or behaviors in a network using clustering algorithms and achieve network-level and cluster-level enhancements. & MAC & Non-RT RIC\\\cline{2-4}
 \hline
   \multirow{3}{*}{Reinforcement Learning}   & Optimal allocation of radio resources in high mobility networks, e.g., UAVs, using reinforcement learning-based control algorithms.  & MAC & Near-RT RIC  \\ \cline{2-4}
 & Optimal configuration of massive MIMO beamforming parameters for performance optimization using reinforcement learning algorithms.  & PHY & RT RIC \\\cline{2-4}
 \hline
 {Federated Learning}   & Real-time spectrum sensing and sharing for dynamic access.   & MAC & Near-RT \\\cline{2-4}
 \hline
   \multirow{3}{*}{Transfer Learning}   & Adaption of pre-trained AI models to a new learning task (e.g., new environment) where limited training data is available using domain adaption algorithms.   & MAC/PHY & Near-/Non-RT RIC \\
 \hline
\end{tabular}
\label{tab:AI}
\end{table*}

\subsection{{Testing Framework Requirements}}
Empowered by openness and intelligence, O-RAN is now gaining its importance and popularity in both industry and academia, and many new xApps/rApps have been developed or are being under development. However, how to comprehensively test these new AI-enabled features (i.e., xApps/rApps) becomes a pressing question. We envision that an automated and AI-enhanced testing (AI testing for AI) platform is of great essence and value to test new O-RAN capabilities. On the one hand, automated testing involves automated setup of the testing environment, automated test execution, and automated generation of testing performance reports. On the other hand, for large search spaces, AI methods can be useful to control the inputs and parameters to the system under test. In particular, we identify the following \textit{requirements} for testing AI-enabled RAN controllers:
\begin{itemize}
    \item Software-defined and modular to enable 
    customization,
    \item Invasive/non-invasive testing during O-RAN operation in isolated or production environment to capture data in relevant operating conditions,
    \item Open test interfaces to enable the development of new test methods and processes,
    \item Test configuration files that enable specifying and reproducing a test,
    \item Support for automated and AI-enhanced testing to assess the operation of AI-enabled cellular radio network controllers under a myriad of channel and contextual conditions (large search spaces), 
    \item {Facilitate the acquisition of data for the training of AI models that generate the test signals. 
    }
    \item Support for multitasking and distributed testing to enable a multi-user testing environment. 
\end{itemize}

\begin{figure*}[!h]
	\centering
	\includegraphics[width=15cm]{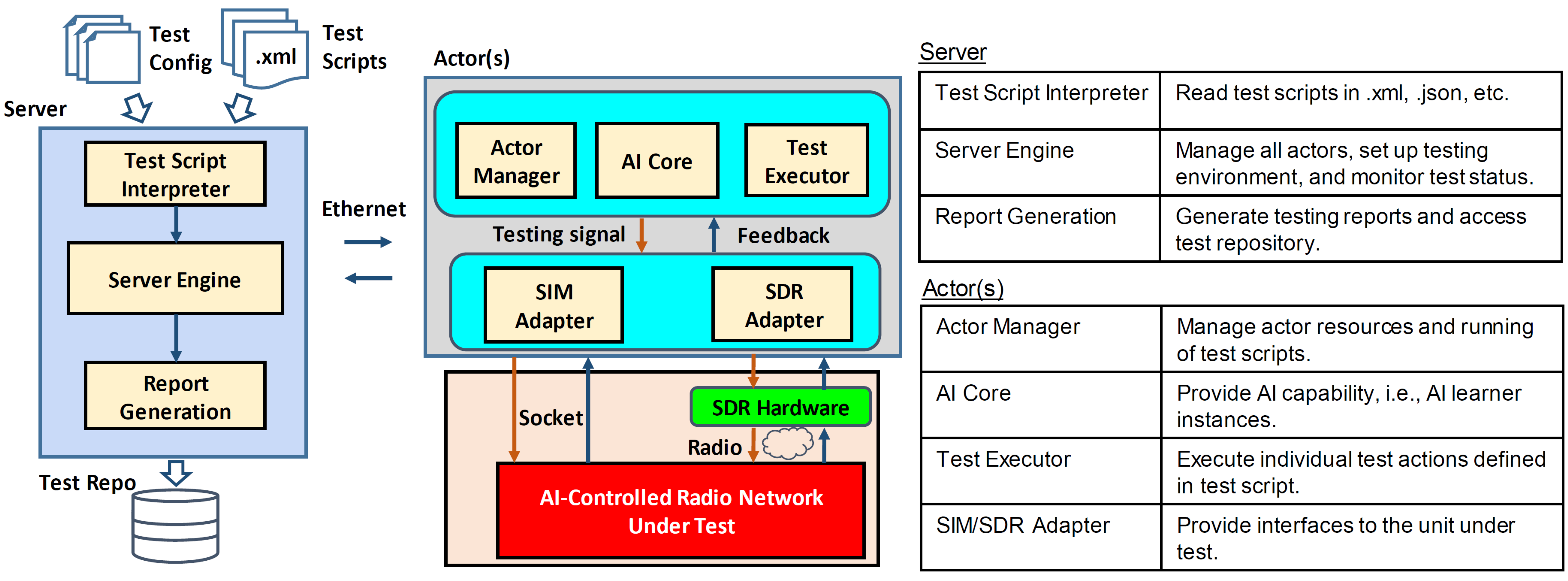}
	\caption{Architecture of the proposed open AI cellular testing framework.}
	\label{Fig:illustrate}
\end{figure*}

\section{Proposed Open AI Cellular Testing Framework}
\label{sec:oaict}

We illustrates the architecture of the proposed framework in Fig.~\ref{Fig:illustrate}, which consists of three major components: \textit{test input}, which includes both test configuration and test script files, \textit{server}, which sets up the testing environment as described in test configuration files and orchestrates the test execution as defined in test scripts, and \textit{actor}, which executes test actions as instructed by the server. 
We detail the three major components of our proposed testing framework as follows:
\begin{itemize}
\item \textit{Test Input}: For each task, this framework takes test configuration and test scripts as inputs. While the test configuration file is used to automatically set up the testing environment, a test script defines the automated test procedure, which consists of a set of test actions to be executed (e.g., an actor sending an Attach Request or responding with feedback). The framework performs the known ``keyword-driven" testing, i.e., each test action is referred to as a ``keyword" and test actions execute sequentially. A test action can be either an ``atomic" or a ``running" action. Only the running actions will be executed sequentially, while a set of sequential atomic actions can be grouped as a new test action. This multi-level granularity of actions 
enables user-defined 
test scripts for various applications (e.g., protocol, functional, performance, or integrating tests). The structure of a test script is illustrated in Fig.~\ref{Fig:testcase}.

\begin{figure}[!h]
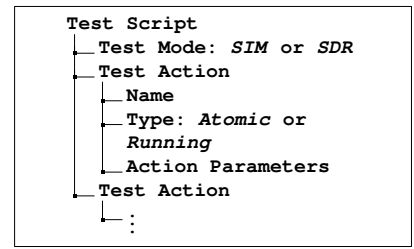

\centering
\resizebox{0.3\textwidth}{!}{
\framebox[7cm]{%
\begin{minipage}{6cm}
  \dirtree{%
  .1 \textbf{Test Script}.
  .2 \textbf{Test Mode: \textit{SIM} or \textit{SDR}}.
  .2 \textbf{Test Action}.
  .3 \textbf{Name}.
  .3 \textbf{Type: \textit{Atomic} or \textit{Running}}.
  .3 \textbf{Action Parameters}.
  .2 \textbf{Test Action}.
  .3
  \vdots.
  }
\end{minipage}}
}
\captionsetup{justification=centering}
\caption{{The structure of a test script.}}
\label{Fig:testcase}
\end{figure}

\item \textit{Server}: The server manages a number of remote testing actors, including maintaining the socket connection to each actor and monitoring its resource usage (e.g. CPU, memory, disk, and SDR hardware) and test status. Before a test starts with one or more test scripts, it first checks the integrity of each test script (i.e., whether all necessary elements are included) and extracts the sequence of test actions. These test actions are then sent to specific test actors, together with the test data if it is included. The server also continuously monitors the status of all running tests which can be displayed to users. Once a test {completes}, 
the server 
collects all test parameters from the actor, generates a report 
of the test results, and 
stores them in a database. 

\item \textit{Actor}: An actor consists of the \textit{actor manager}, \textit{AI core}, and \textit{test executor}, as well as two \textit{adapters} to interface with the unit under test (i.e., a cellular radio network controller). The actor manager is able to monitor the health of the actor (e.g., resource utilization and hardware check) and configure the testing environment for a specific task. The actor also includes a number of AI cores which implement AI-enabled testing algorithms to facilitate the testing with the help of AI. 
The test executor is responsible for running sequential test actions as instructed by the server and {for} monitoring their {statuses} which are reported to the server. Each actor can interact with xApps/rApps under test through either the \textit{SIM Adapter} or the \textit{SDR Adapter}. The SIM Adapter acts as a testing xApp/rApp which can be deployed in a non-RT, near-RT, or future RT RIC to send testing data and receive responses from the RAN through sockets. 
The SDR Adapter acts as a UE which leverages an SDR-based software suite (e.g., srsRAN) to send radio testing signals to the RAN. 
\end{itemize}

\begin{figure}[!h]
	\centering
	\includegraphics[width=8cm]{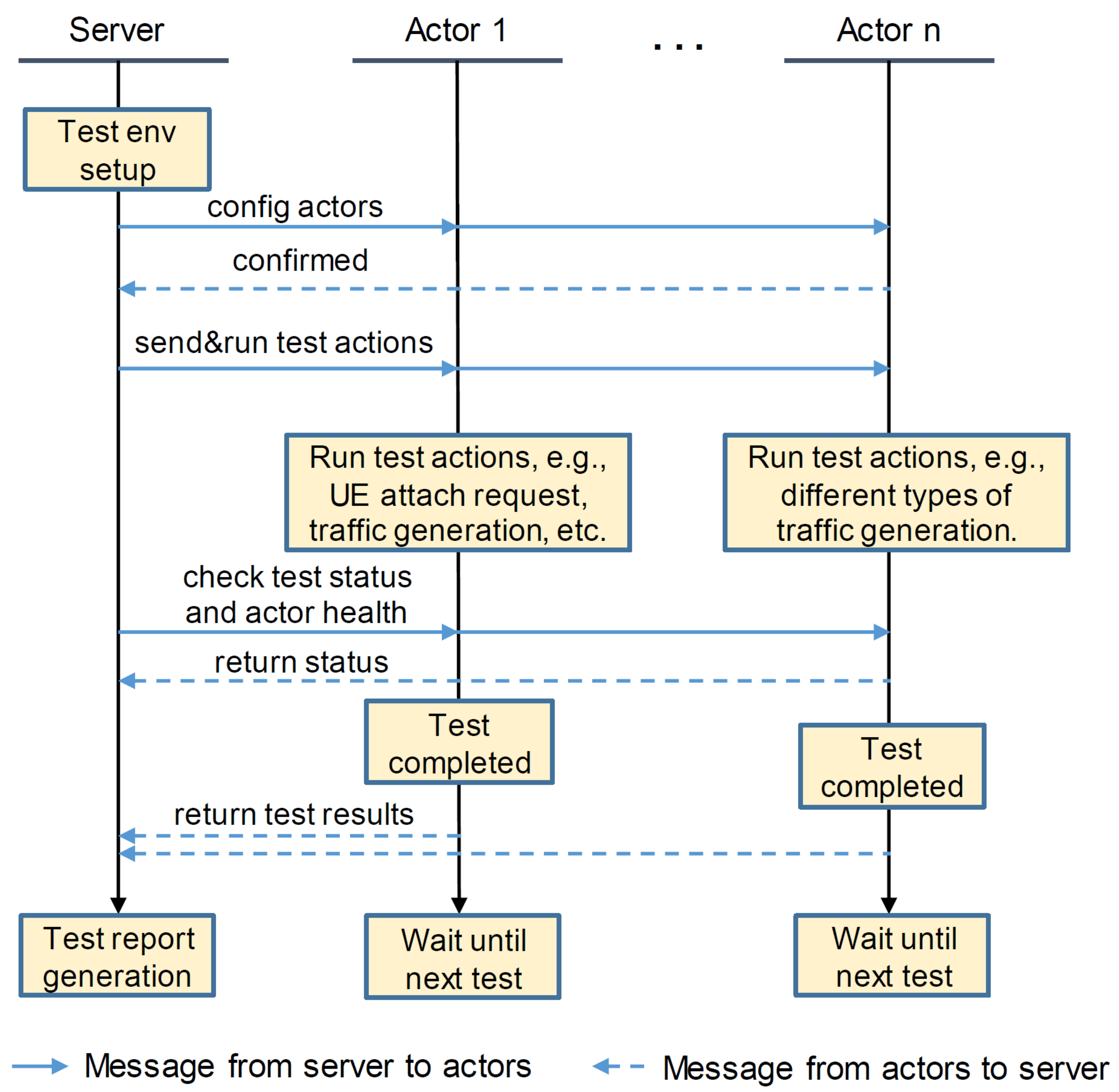}
	\caption{A sample workflow for testing schedulers in O-RAN.}
	\label{Fig:workflow}
\end{figure}

Whereas most tests can be performed {on} a single machine, multiple actors deployed on different machines in the network can be involved in the same test to support distributed testing. For example, when testing an AI-enabled scheduler for radio resource allocation, a number of distributed actors can be configured to request various radio resources with different quality of service (QoS) requirements or priority levels. The number of actors and their associated Internet protocol addresses 
are defined in the test configuration file that the server can use to set up the testing environment. Fig.~\ref{Fig:workflow} shows a sample workflow for testing schedulers in O-RAN. 


After a test is done, the framework automatically generates a test report which details the test setup and {result} of each individual test {action} (fail or {pass}), and summarizes various performance metrics of the network under test. In addition to the overall success rate of the test actions, the proposed framework can also summarize and visualize 
{network KPIs},  such as data rate, latency, and packet loss. 

\section{Enabling Technologies -- AI Testing Methods}
\label{sec:capabilities}


With the use of various AI models for wireless networks, measuring the performance of the deployed models is a necessity for quality and security assurance (e.g., service delivery, response time, resource allocation optimization, vulnerability assessment, and mitigation). This is due to the fact that the effectiveness of AI systems in their decision making processes mainly depends on the quality of the training data which might not cover all possible practical conditions. To bridge this gap, the proposed {open AI cellular testing} framework {introduces} AI (i.e., AI core in the actor) in the test automation system 
to generate dynamic test actions and autonomously explore the decision making capabilities of AI-controlled RANs. The generation of test actions will be guided by specific AI testing methods (e.g., Fuzz testing, reinforcement learning, and adversarial learning) as defined in the test script, and all details (i.e., parameters, states, and actions) of this automated exploratory process will be recorded as part of the output report which further 
allows to replicate experiments for regression testing. Here, we highlight the following enabling technologies {for} AI testing to be integrated into the framework:
\begin{itemize}
\nointerlineskip
     \item {\textit{
     Sensitivity Analysis}: The 
     sensitivity analysis assesses the O-RAN processes with controlled 
     changes in the operating conditions using expert knowledge, More precisely, given a parameter space, each test varies only one or a few variables at a time. This may trigger system adaptation and the response of the system is compared to the expected 
     system response. 
     Some environmental effects should trigger changes {while} others should not, and the goal is to observe the sensitivity of the xApps/rApps at different levels or perturbation. 
     Such testing may allow to interpret the behavior of the black box AI models used in the xApps or rApps and provide insights on the stability of the system. 
     It can be considered as a vulnerability analysis and can serve to evaluate the security and reliability of the network.}
    \item \textit{AI Fuzzing}: Fuzz testing or Fuzzing has been widely used in automated software testing~\cite{takanen2018fuzzing} which autonomously generates input data with perturbation 
    to find software defects and faults (e.g., an unhandled exception or memory leak). Coupling AI methods (e.g., genetic algorithms) with Fuzzing, i.e. AI-Fuzzing (AIF), 
    has great potential to effectively and intelligently evaluate the performance of AI-controlled systems. It works on the principles of dynamic test generation to explore the large input space and decision boundaries of the system. For example, consider the multi-class support vector machine (SVM)-based demodulator used in carrier frequency offset (CFO)-impaired systems~\cite{Siyari_CFO_SVM_JSAC_2019}. For each CFO value, the CFO-impaired received data symbols are used to train the SVM classifier to learn the optimal decision boundaries for each quadrature amplitude modulation (QAM) constellation point. To test this module using the proposed testing framework, the test parameters such as initial CFO values, the number of data symbols, and Fuzzing parameters are specified in the test script and parsed by the test script interpreter. Instead of only using initial CFO values, AIF 
    generates a sequence of test signals with different impaired CFO values to explore the decision boundaries of the trained SVM classifier. 
    
    \item \textit{Adversarial Learning}: For the testing of AI-controllers for physical layer modules, adversarial learning~\cite{huang2011adversarial} enables automated generation of test signals which are able to capture similar characteristics of real-world channel conditions (e.g., I/Q imbalance or non-Gaussian interference). This offers the capability of intelligent stress testing of a wireless cellular system using inputs which may violate the statistical assumption of AI methods under test. 
    Moreover, motivated by the recent success of {generative adversarial networks}, it is also possible to enable AI-testing of system security against various cyber-attacks, for instance, spoofing attacks which mimic {channel state information} and RF fingerprints of legitimate radio signals, and denial of service attacks which flood the target network with a significant amount of traffic. 
    
    \item \textit{Reinforcement Learning}: 
    Reinforcement learning models {learn} how to take actions, or adjustments, in an 
    environment to maximize a predefined cumulative reward, e.g., the negative QoS requirement in case of a scheduler, so that its vulnerability surface can be effectively explored. For each traffic under test, the reinforcement learning-based procedure will parse the response of the system under test (e.g., AI-enabled scheduler) as a reward and decide how to allocate radio resources in the next step. This will enable the AI testing platform to learn during the process of testing (under several test environments), and optimize its testing strategies on the fly so that the performance of the AI-enabled scheduler can be exhaustively analyzed.
\end{itemize}
It is worth noting that the above mentioned AI modules can be designed in a way that each of them can be used for standalone testing. Combining them can make the testing framework more powerful. Moreover, instead of testing the AI-controlled RAN as a black box, this framework allows to add white box embedding components in the RAN under test to monitor and record all detailed behaviors when testing inputs are applied.
New learning capability can be easily added to the AI library. 

\section{{Challenges and} Research Opportunities}
\label{sec:research}

With the increasing popularity of O-RAN, 
it is expected that many more advanced AI-controlled functions---xApps and rApps---will be developed 
by the community. The proposed testing framework enables {the} comprehensive testing of these functions in an automated, distributed, and AI-enabled manner. 
{This opens up a number of research opportunities to facilitate the 
prototyping, deployment, and operation of Next-G 
networks.} Here we briefly discuss the critical 
research needed and 
enabled by the proposed testing framework and its limitations that will be addressed in our future work. 

\begin{itemize}
    
    \item \textit{
    AI Security}: The benefits of using AI models are obvious in wireless networks: they enable faster analysis of large scale data (e.g., traffic or RF signals) and can make 
    better (optimal or near-optimal) decisions than a  human. 
    These benefits come from the unique value of AI models which learn from data. However, this also happens to make them more vulnerable and less trustworthy from a security perspective. When an unexpected behavior or decision is observed, an AI model is not as easily fixed by hand-editing as are traditional algorithms or formulas. This problem might be addressable today if we had explainable AI, but coming up with general approaches for explaining AI solutions has alluded researchers for decades. 
    Given the high value of commercial cellular wireless networks, there is a growing incentive for malicious attackers to explore and exploit possible vulnerabilities in AI models. There are three types of attacks that adversaries could launch: model evasion attack, model poisoning attack, and confidentiality attack. The model evasion attack aims to feed adversarial samples which are carefully perturbed into a ML model (e.g., anomaly detection) so that a wrong decision will be made. While great harm can be caused by model evasion attacks, they do not alter the behavior of the trained model for future inputs. On the contrary, the poisoning attack aims to intentionally and maliciously adjust decision boundaries of a model so that it always misclassifies specific 
    inputs. This can be typically done during the training stage of the model by feeding ``poisoned" data. Lastly, the confidentiality attack aims to replicate a model and/or reveal sensitive data used to train the model, both of which are protected by intellectual property rights. This can be typically done by recursively querying the model with different input. Using the proposed testing framework, new research in AI vulnerabilities are enabled to improve the understanding of behaviors of AI models for different inputs or to evaluate the effectiveness of new attack and defense mechanisms, among others.

    \item \textit{
    AI Testing}: Before the emergence of advanced AI algorithms that are more interpretable and trustworthy, it is very difficult to fully understand behaviors of current data-driven AI models without the help of AI. It has been demonstrated that a comprehensive test is one of the most effective ways to expose vulnerabilities and potentially help improve the trustworthiness of AI models. The proposed testing framework supports the deployment of both offline and online AI testing methods and has the potential to be extended to test various performances including correctness, robustness, and efficiency. The correctness measures the probability that the AI model makes correct decisions. The robustness measures the performance of the AI model in the presence of invalid inputs or valid inputs but perturbed with noise, which represent different 
    environmental conditions. The efficiency measures the speed of an AI model to make effective decisions after an input or condition is presented. While the proposed testing framework integrates some existing AI testing methods, such as AIF, we expect a number of new AI testing methods {to} be developed in the near future. For example, testing methods that leverage advanced generative models are able to generate more realistic data inputs that locate around decision boundaries of an AI model to explore its correctness and robustness performances. 
    {A major challenge is the availability of training data to train the AI testing models. 
    Similar to industry and academia using RAN simulators or emulators to develop the 
    xApps or rApps and collect performance data, the testing framework needs to facilitate the integration of 
    tools 
    for the generation and collection of data under various experimental conditions. In addition, the system overhead pertaining to training AI models is another challenge for an AI testing framework since it 
    needs to generate test signals while performing the testing.}
    
    \item \textit{
    Coordinated Testing}: Wireless networks could become much more vulnerable when multiple malicious nodes launch coordinated attacks, e.g., denial of service attacks. Such coordinated attacks are {hard to detect} using existing defense methods which are mainly developed for detecting single node attacks, and they become more challenging particularly when attack data samples are unbalanced, high dimensional, and noisy. This requires {a better understanding of} how a wireless network {behaves} in the presence of such coordinated attacks. The proposed testing framework can be extended to support coordinated {execution} of test actions {from} a number of distributed actors. This can be done by allowing message communications among different actors. The coordinated testing capability will enable the {prototyping} and evaluation of various cooperative attack and defense mechanisms targeting O-RAN based wireless networks. 
    
\end{itemize}

\section{Conclusions}
\label{sec:conclusions}

AI {is the enabling technology} to achieve intelligent Next-G wireless networks. However, the performance of AI models heavily depends on the quality and quantity of their training data, as well as their generalization capability. There is a lack of a general testing framework to comprehensively test the performance of AI models deployed in the RAN. This article presents an automated, distributed and AI-enabled testing framework which has the potential to fully evaluate the performance of AI models in terms of their decision-making performance, vulnerability, and security {in the context of O-RAN}. {The proposed framework} leverages AI to generate testing signals in an automated and intelligent manner so that the decision space of the AI models used in O-RAN can be explored. Under the umbrella of the proposed testing framework, this article also discusses enabling techniques for AI testing and important research opportunities 
in the field of AI, wireless communications, cyber security, and coordinated testing. 

\bibliographystyle{IEEEtran}
\bibliography{./bibtex/bib/vuk, ./bibtex/bib/bo}

\begin{thebibliography}{10}
\providecommand{\url}[1]{#1}
\csname url@samestyle\endcsname
\providecommand{\newblock}{\relax}
\providecommand{\bibinfo}[2]{#2}
\providecommand{\BIBentrySTDinterwordspacing}{\spaceskip=0pt\relax}
\providecommand{\BIBentryALTinterwordstretchfactor}{4}
\providecommand{\BIBentryALTinterwordspacing}{\spaceskip=\fontdimen2\font plus
\BIBentryALTinterwordstretchfactor\fontdimen3\font minus
  \fontdimen4\font\relax}
\providecommand{\BIBforeignlanguage}[2]{{%
\expandafter\ifx\csname l@#1\endcsname\relax
\typeout{** WARNING: IEEEtran.bst: No hyphenation pattern has been}%
\typeout{** loaded for the language `#1'. Using the pattern for}%
\typeout{** the default language instead.}%
\else
\language=\csname l@#1\endcsname
\fi
#2}}
\providecommand{\BIBdecl}{\relax}
\BIBdecl

\bibitem{wang2020artificial}
C.-X. Wang, M.~Di~Renzo, S.~Stanczak, S.~Wang, and E.~G. Larsson, ``Artificial
  intelligence enabled wireless networking for {5G} and beyond: Recent advances
  and future challenges,'' \emph{IEEE Wireless Communications}, vol.~27, no.~1,
  pp. 16--23, 2020.

\bibitem{shafin2020artificial}
R.~Shafin, L.~Liu, V.~Chandrasekhar, H.~Chen, J.~Reed, and J.~C. Zhang,
  ``Artificial intelligence-enabled cellular networks: A critical path to
  beyond-{5G} and {6G},'' \emph{IEEE Wireless Communications}, vol.~27, no.~2,
  pp. 212--217, 2020.

\bibitem{etsi}
ETSI, ``{Experiential Networked Intelligence (ENI): ENI Definition of
  Categories for AI Application to Networks },'' March 2021.

\bibitem{ORANwhitepaper}
{Open-RAN Alliance}, ``{O-RAN}: towards an open and smart {RAN},'' \emph{White
  Paper}, pp. 1--19, 2018.

\bibitem{oaic_msstate}
{Open AI Cellular (OAIC)}, ``{Prototyping Artificial Intelligence-Enabled
  Control and Testing Systems for Cellular Communications Research},''
  \url{https://www.openaicellular.org/}, accessed: 2022-08-28.

\bibitem{ORAN_Testing}
{O-RAN Alliance Test and Integration Focus Group}, ``{End-to-end Test
  Specification},'' {Technical Specification O-RAN.TIFG.E2E-Test.0-v02.00},
  2021.

\bibitem{etsi_testing}
ETSI, ``{Artificial Intelligence (AI) in Test Systems, Testing AI Models and
  ETSI GANA Model's Cognitive Decision Elements (DEs) via a Generic Test
  Framework for Testing GANA Multi-Layer Autonomics \& their AI Algorithms for
  Closed-Loop Network Automation},'' {White Paper No.5}, March 2020.

\bibitem{mina17}
M.~{Labib}, V.~{Marojevic}, J.~H. {Reed}, and A.~I. {Zaghloul}, ``Enhancing the
  robustness of {LTE} systems: Analysis and evolution of the cell selection
  process,'' \emph{IEEE Communications Magazine}, vol.~55, no.~2, pp. 208--215,
  February 2017.

\bibitem{sean17}
V.~{Marojevic}, R.~M. {Rao}, S.~{Ha}, and J.~H. {Reed}, ``Performance analysis
  of a mission-critical portable {LTE} system in targeted {RF} interference,''
  in \emph{2017 IEEE 86th Vehicular Technology Conference (VTC-Fall)}, Sep.
  2017, pp. 1--6.

\bibitem{Byrd2020}
T.~Byrd, V.~Marojevic, and R.~P. Jover, ``{CSAI}: Open-source cellular radio
  access network security analysis instrument,'' in \emph{2020 IEEE 91st
  Vehicular Technology Conference (VTC2020-Spring)}, 2020, pp. 1--5.

\bibitem{prochecker}
I.~Karim, S.~R. Hussain, and E.~Bertino, ``Prochecker: An automated security
  and privacy analysis framework for {4G LTE} protocol implementations,'' in
  \emph{2021 IEEE 41st International Conference on Distributed Computing
  Systems (ICDCS)}, 2021, pp. 773--785.

\bibitem{ORANTechRep20}
{Open-RAN Alliance}, ``{O-RAN Working Group 2 AI/ML workflow description and
  requirements},'' \emph{Technical Report O-RAN.WG2.AIML-v01.02.02}, pp. 1--51,
  2020.

\bibitem{takanen2018fuzzing}
A.~Takanen, J.~D. Demott, C.~Miller, and A.~Kettunen, \emph{Fuzzing for
  software security testing and quality assurance}.\hskip 1em plus 0.5em minus
  0.4em\relax Artech House, 2018.

\bibitem{Siyari_CFO_SVM_JSAC_2019}
P.~{Siyari}, H.~{Rahbari}, and M.~{Krunz}, ``{Lightweight Machine Learning for
  Efficient Frequency-Offset-Aware Demodulation},'' \emph{IEEE Journal on
  Selected Areas in Communications}, vol.~37, no.~11, pp. 2544--2558, Nov 2019.

\bibitem{huang2011adversarial}
L.~Huang, A.~D. Joseph, B.~Nelson, B.~I. Rubinstein, and J.~Tygar,
  ``Adversarial machine learning,'' in \emph{Proceedings of the 4th ACM
  workshop on Security and artificial intelligence}.\hskip 1em plus 0.5em minus
  0.4em\relax ACM, 2011, pp. 43--58.

\end{thebibliography}

\ifCLASSOPTIONcaptionsoff
  \newpage
\fi

%




\end{document}